\documentclass[review]{elsarticle}
\graphicspath{ {} }
\usepackage{hyperref}
\usepackage{amssymb}
\usepackage[table,xcdraw]{xcolor}
\usepackage{colortbl}
\usepackage{float}
\usepackage{verbatim} 
\usepackage{apalike}
\usepackage[utf8]{inputenc}
\restylefloat{figure}
\restylefloat{table}

\usepackage{algorithm2e}
\usepackage{graphicx}
\usepackage{subcaption}
\usepackage{glossaries}
\usepackage{amsmath}

\journal{Expert Systems with Applications}

\biboptions{authoryear}

\begin{document}

\begin{titlepage}
\begin{center}
\vspace*{1cm}

\textbf{ \large A Novel Approach to Fair Power Allocation for NOMA in Visible Light Communication}

\vspace{1.5cm}

Serkan VELA$^{a,b}$ (serkanvela@ktu.edu.tr), Gokce HACIOGLU$^{b,*}$ (gokcehacioglu@ktu.edu.tr)  \\

\hspace{10pt}

\begin{flushleft}
\small  
$^a$ Karadeniz Technical University, Of Technology of Faculty, Electronics and Communication Engineering, Trabzon/TURKEY \\
$^b$ Karadeniz Technical University, Engineering Faculty, Electrical and Electronics Engineering, Trabzon/TURKEY \\

\vspace{1cm}
\textbf{Corresponding Author:} \\
Gokce HACIOGLU \\
Karadeniz Technical University, Engineering Faculty, Electrical and Electronics Engineering, Trabzon/TURKEY \\
Tel:+90 (462) 377-2077 \\
Email: gokcehacioglu@ktu.edu.tr

\end{flushleft}        
\end{center}
\end{titlepage}

\title{A Novel Approach to Fair Power Allocation for NOMA in Visible Light Communication}

\author[label1,label2]{Serkan VELA}
\ead{serkanvela@ktu.edu.tr}

\author[label2]{Gokce HACIOGLU \corref{cor1}}
\ead{gokcehacioglu@ktu.edu.tr}

\cortext[cor1]{Corresponding author.}
\address[label1]{Karadeniz Technical University, Of Technology of Faculty, Electronics and Communication Engineering, Trabzon/TURKEY}
\address[label2]{Karadeniz Technical University, Engineering Faculty, Electrical and Electronics Engineering, Trabzon/TURKEY}

\begin{abstract}
The demand for high-bandwidth wireless data transmission has prompted exploration beyond the confines of Radio Frequency (RF) communication. Visible Light Communication (VLC), with its advantages such as high bandwidth, license-free spectrum, immunity to electromagnetic interference, and low power consumption, emerges as a compelling alternative or complement to RF systems. Particularly in indoor scenarios, VLC systems leverage existing lighting infrastructure for high-speed data transmission.

To meet the data rate demands of 5G and beyond, Non-Orthogonal Multiple Access (NOMA) presents itself as a promising technology. NOMA involves superimposing signals from multiple users in the power domain, offering low transmission delay, increased spectrum efficiency, and higher total data rates. However, ensuring fair distribution of the sum rate among users poses a critical challenge in NOMA, necessitating the use of optimization algorithms.

In this paper, we introduce a new approach called Empirical Fair Optical Power Allocation (EFOPA) aimed at reducing the computational complexity associated with resource allocation in NOMA. In this approach, we integrate NOMA with VLC systems. Initially, we employ the Artificial Bee Colony (ABC) optimization algorithm to formulate an offline resource allocation plan that can accommodate various channel conditions. Subsequently, we use the results from ABC to derive a simplified power allocation equation, ensuring fair resource allocation among users. This approach assigns power to NOMA users based on a simplified equation derived from ABC outcomes, facilitating the fair distribution of resources among users. Importantly, EFOPA maintains fairness even in the presence of variations in illumination intensity and channel gains.

The comparative evaluations against well-known power allocation methods, including Gain Ratio Power Allocation (GRPA), Normalized Gain Difference Power Allocation (NGDPA), and Orthogonal Multiple Access (OMA), underscore the superiority of EFOPA. EFOPA not only outperforms existing methods in terms of fairness but also significantly reduces computational complexity by utilizing an empirical equation. The results demonstrate that EFOPA achieves similar or even better sum rates in nearly 50\% of the time compared to GRPA, while outperforming OMA and NGDPA under various channel conditions. In numerical comparisons, EFOPA consistently outperforms OMA in nearly 98\% of all possible channel conditions and surpasses NGDPA in more than 90\% of channel conditions. The method's efficacy is further highlighted, showcasing EFOPA as a robust and efficient solution for fair power allocation in NOMA-VLC systems. 
\end{abstract}


\section{Introduction}
\label{introduction}
The demand for high-bandwidth wireless data transmission is increasing day by day. The Radio Frequency (RF) spectrum cannot fulfill the high data rate demand cost-effectively. Visible Light Communication (VLC) emerges as a promising alternative or companion to RF systems due to its high bandwidth, license-free spectrum, immunity to electromagnetic interference, and low power consumption \citep{ahsan2017esim,oyewobi2022visible,memedi2020vehicular,abuella2021hybrid}. In indoor scenarios, VLC systems utilize existing lighting infrastructure to provide high-speed data transmission \citep{pathak2015visible,armstrong2013visible,li202140,yu2021visible}. \citep{singh2019power} compares RF and VLC links in terms of symbol-error-rate (SER) performance and power savings, finding that VLC offers better SER performance and significant power savings compared to RF in indoor environments. Additionally, \citep{abuella2021hybrid} mentions that VLC offers a broader and unused frequency spectrum compared to RF, minimizing congestion.

In VLC systems, data is modulated by the intensity of the light. Therefore, transmitted signals should always be positively valued \citep{gong2015power,kashef2014benefits,shen2016user,shen2016rate}. Although a wide unlicensed band is available for VLC systems, the switching frequency of the Light Emitting Diodes (LED) determines the communication speed. The switching frequencies of blue-phosphor layer LEDs used in lighting are a few MHz \citep{komine2004fundamental,dou2023achieving,vitasek2019comparison}. On the other hand, RGB LEDs can be switched with tens of MHz \citep{medina2015led}.

Non-Orthogonal Multiple Access (NOMA) was suggested as a promising candidate to increase the total data rate of multiple users, especially for 5G wireless networks \citep{benjebbour2013concept,liu2017nonorthogonal,wei2016survey,islam2016power,dai2015non,ding2017application}. In NOMA, signals belonging to more than one user are superimposed and transmitted in the power domain without separating in the time or frequency domain \citep{timotheou2015fairness,sadia2018performance}. Signals at receivers are obtained by the successive interference cancellation (SIC) method \citep{manglayev2017noma,tao2018performance,jha2022noma,iraqi2021power}. Each user can use the same band at the same time. In this respect, NOMA has low transmission delay, increased spectrum efficiency, and a higher total data rate \citep{islam2016power,karim2023modeling,nguyen2020survey}.

To take advantage of NOMA efficiently, various studies have been carried out. The Optimum Power Allocation (OPA) algorithm, which takes into account the data rate of users relative to each other (fairness), has been proposed in \citep{manglayev2016optimum}. A power allocation algorithm for energy and spectral efficiency using enhanced particle swarm optimization (PSO) has also been proposed in \citep{xiao2018improved}. \citep{chikezie2022power,abuajwa2022resource,li2020proportional} investigates fair power allocation in 5G networks and proposes improved fair power allocation methods that outperform traditional schemes. However, these algorithms were proposed for the RF usage of NOMA. The usage of NOMA in VLC should be studied separately as a particular case. In a cornerstone study \citep{marshoud2015non}, the NOMA method was suggested to increase the data rate in VLC. Various power allocation algorithms such as gain ratio power allocation (GRPA) \citep{marshoud2015non}, normalized GRPA \citep{chen2017performance} and normalized gain difference power allocation (NGDPA) \citep{chen2017performance,yin2016performance,zhang2016user} have been proposed to increase the overall data rate. GRPA strategies initially developed for RF channels are ineffective in VLC systems. In \citep{yang2021power}, an optimization algorithm was proposed to maximize the sum rate and compare it with GRPA. The algorithm has low complexity for only specific cases, as they admit. Fairness and illumination are not considered in these studies. The lighting level and fairness should be considered to ensure comfort and Quality of Service (QoS) while allocating the power of the transmitter LEDs to the users in VLC systems. The decoding order power allocation (DOPA) method, as introduced by \citep{alqahtani2022decoding}, employs the Golden Section Search and Parabolic Interpolation (GSSPI) algorithm to identify optimal power allocations and then stores this information in a lookup table. However, it may have limitations in adapting effectively to environments that differ from the specifically simulated one, particularly when considering mobile users with varying channel gains.

In the study \citep{tahira2019optimization}, where the receivers’ positions are fixed, the total data rate and bit error rate (BER) under a particular light intensity constraint are optimized with a convex solver and compared with the static power allocation. However, the data rate of the users and fairness were not considered. In \citep{tao2018power}, the GRPA method was used with variable on-off keying (VOOK) for fixed receivers to allocate power in dimmable NOMA-VLC systems. VOOK is an inefficient method as the total capacity decreases in proportion to the brightness. In \citep{eltokhey2021power}, at the cost of increased complexity, a particle swarm optimization (PSO) based power allocation optimization algorithm for the NOMA VLC multi-cell network is proposed. The level of fairness obtained with this method is less than 0.6 and is not ideal for QoS.

In the context of this study, fairness is not merely a consideration but is embraced as a central objective function. The optimization endeavors are executed through the utilization of the Artificial Bee Colony (ABC) algorithm, inspired by the foraging methods observed in honey bees \citep{karaboga2007artificial}. In this research, the pursuit of maximum fairness is metaphorically modeled as the collective quest for a food source by the bees, with the constraints inherent to Visible Light Communication (VLC) serving as the pathways available to these diligent foragers. Consequently, we introduce the Empirical Fair Optical Power Allocation (EFOPA) method, which not only ensures the attainment of maximum fairness but also boasts low computational complexity and supports dimming capabilities. EFOPA is founded on a straightforward equation for fair optical power allocation. Noteworthy is the absence, to the best of our knowledge, of a comparable simplified fair optical power allocation equation in existing literature, catering comprehensively to both static and dynamic receivers while guaranteeing maximum fairness and customizable illumination levels.

The system model, VLC channel, and NOMA are discussed in section (\ref{System Model, VLC Channel and PD-NOMA}). A brief information about the ABC algorithm and its use for the proposed method is given in section (\ref{Optimization With ABC Algorithm}). The proposed model (EFOPA) is presented in section (\ref{Numerical Results and Discussions}). The performance of EFOPA is compared with the GRPA and NGDPA methods in section (\ref{results}), and the paper is concluded in section (\ref{Conclusion}).

\section{System Model, VLC Channel, and PD-NOMA}
\label{System Model, VLC Channel and PD-NOMA}
\begin{figure}
\includegraphics[scale=0.24]{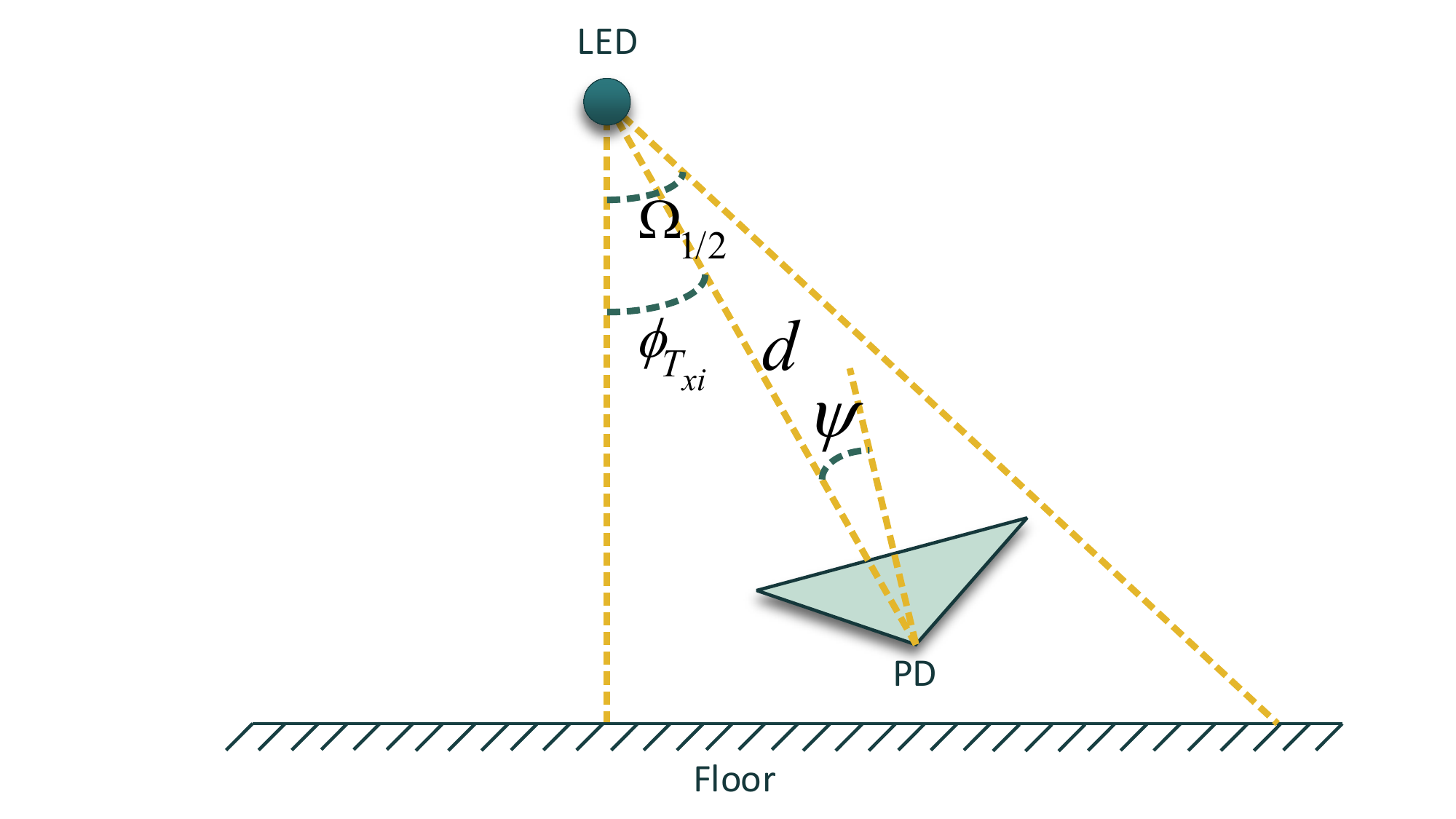}
\centering
\caption{LoS Model for VLC Channel}
\label{fig:VLC}
\end{figure}

In this section, we examine the system model, VLC channel characteristics, and the integration of Power Domain-NOMA (PD-NOMA) in an indoor visible light communication setting within an office room. The LED transmitter is strategically positioned on the ceiling of the room, facilitating data transmission to users equipped with a single photo-detector (PD) within the field of view (FoV). The VLC channel prominently features a robust Line of Sight (LoS) component. To illustrate, the most attenuated Non-Line of Sight (NLoS) component in the VLC channel is $7dB$ weaker than the least powerful LoS component, as demonstrated in \citep{zeng2009high}. Consequently, our analysis exclusively considers the LoS component. Figure \ref{fig:VLC} visually represents the LoS model for the VLC channel.

The VLC channel gain between the transmitter and the $k^{th}$ user is denoted by $(h_k)$ and is expressed below.
\begin{equation}\label{hk}
h_k=\frac{A_kR\left(\phi_{k}\right)}{d^2}T_s\left(\psi\right)g\left(\psi\right)\cos{\left(\psi\right)}, \ \ \ \ \ \ \ \ \ \ \ \ 0\le\psi\le\phi
\end{equation}
$A_k$ is the PD area of $k^{th}$ user, $d$ is the distance between transmitter and receiver, $\phi_{k}$ is the angle of irradiance with respect to the transmitter perpendicular axis, $\psi$ is the angle of incidence with respect to the receiver axis, $\phi$ is the FOV of the receiver, $T_s\left(\psi\right)$ is the gain of optical filter, and $g\left(\psi\right)$ represents the gain of optical concentrator which is given by (\ref{gpsi}).
\begin{equation}\label{gpsi}
g\left(\psi\right)=\frac{n^2}{{sin}^2\phi}
\end{equation}
Where; $n$ is refractive index and $g\left(\psi\right)=0$ for $\psi>\phi$. $R\left(\phi_{k}\right)$ is the Lambertian radiant intensity of the LED which is expressed as the following.
\begin{equation}
R\left(\phi_{k}\right)=\frac{k_l+1}{2\pi}{cos}^{k_l}\left(\phi_{k}\right)
\end{equation}
Where; $k_l$ is the order of Lambertian emission.
\begin{equation}
k_l=\frac{-ln2}{ln{\left(cos{\left(\mathrm{\Omega}_{1/2}\right)}\right)}}
\end{equation}
$\mathrm{\Omega}_{1/2}$ is the transmitter semi-angle at half power.

\begin{figure}
    \centering
    \includegraphics[scale=0.79]{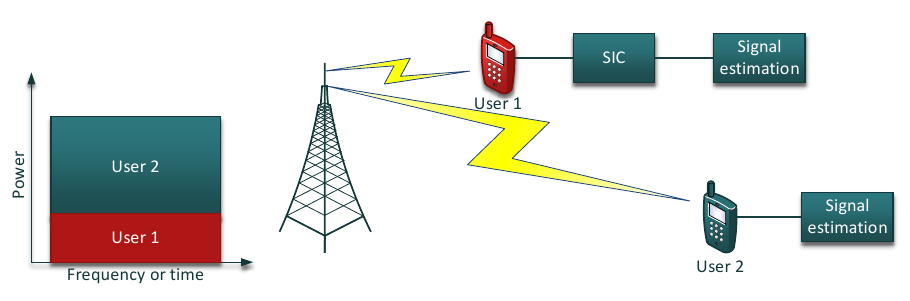}
    \caption{NOMA demo system}
    \label{fig:noma}
\end{figure}

Power Domain Non-Orthogonal Multiple Access (PD-NOMA) represents a multiple access technique wherein users can concurrently utilize the complete time and frequency domains, while their separation occurs in the power domain. In PD-NOMA, diverse power levels are assigned to users based on their respective channel gains, as detailed in \citep{cai2017modulation}. This allocation of power plays a pivotal role in influencing parameters such as data rate, fairness, and illumination level, as highlighted in \citep{serkan}. Therefore, according to channel information theory, lower power allocation should be assigned to the user with a better channel for the maximum utilization of channel capacity \citep{cai2017modulation}. In Figure \ref{fig:noma}, a demo system is depicted with two mobile users using PD-NOMA. 
Let $s_1, s_2$ be the modulated signals for users, and $p_1, p_2$ be the power allocation coefficients. The signal transmitted from the base station is given by $x=\sqrt{p_1} s_1+\sqrt{p_2} s_2$. The signal received at the $k$-th receiver, denoted as $y_k$ for receiver number $k$, is expressed as follows:
\begin{equation}
y_k=h_k x+n_k
\end{equation}
where $h_k$ and $n_k$ are the channel coefficient and noise for the $k$-th user, respectively. Assuming that the channel of User 2 is worse than that of User 1, according to the information theory principle \citep{cai2017modulation}, the message of the user with the worse channel is brought to a higher power level. In this case, the power level of User 2 will be above that of User 1, as shown in Figure \ref{fig:noma}. These two message signals exit the transmitter collectively and then propagate through the channels to reach the users. In the signal received by User 2, the stronger signal is the message signal of User 2. User 2 directly decodes the signal received. User 1, however, in the received collective signal, contains the message of User 2 with the higher power. Initially, User 1 eliminates this signal using the Successive Interference Cancellation (SIC) method. Then, User 1 decodes its own signal. In PD-NOMA, every user except the one with the worst channel must perform SIC. Subsequent to the power allocation procedures, the signals corresponding to each user are superimposed as follows.
\begin{equation}
x=\sum_{k=1}^{K}{\sqrt{p_k}s_k}+A
\end{equation}
 $K$ represents the total number of users. The allocated power and symbol for the $k^{th}$ user is denoted by $p_k$ and $s_k$ respectively. A DC offset $A$ is added to the superimposed form of the user signals to obtain the uni-polar positive valued signal $x$.
\begin{equation}
A\geq \sum_{k=1}^{K}{\sqrt{p_k}s_k}
\end{equation}
The illumination is dependent on the levels of the signal $x$ which is transmitted by the LED. It is assumed that $\sum s_k$ is bipolar and takes normalized values from $-1$ to $1$ ($s_k\in[-1,1]$). The minimum DC offset value ($A$) should be equal to $\sum \sqrt{p_k}$ to ensure non-negativity of $x$. The transmitted signal $x$ passes through the VLC channels and reaches the receivers. The $y_k$ denotes the received signal by $k^{th}$ user is defined in (\ref{yk}).
\begin{equation}
\label{yk}
y_k=h_kx+w
\end{equation}
Where; $w$ represents the Additive White Gaussian Noise (AWGN). After the removal of the DC offset, users should perform the SIC method \citep{manglayev2017noma,tao2018performance}. Applying the well-known SIC algorithm, user $k$ can eliminate multi-user interference caused by other users that have relatively weak channels. In this way, the lower bound expression of VLC channel capacity $R_k$ for the $k^{th}$ user is \citep{hsiao2019energy}
\begin{equation}\label{lowerbound}
R_{k} \triangleq \frac{B_{k}}{2} \log _{2}\left[1+\frac{2 h_{k}^{2} p_{k}}{\pi e\left(h_{k}^{2} \sum_{l=k+1}^{K} p_{l}+\sigma^{2}\right)}\right]
\end{equation}
$B_k$ is the transceiver bandwidth of the $k^{th}$ user and $\sigma^2$ is the variance of AWGN.
\section{Power Allocation with ABC Algorithm}\label{Optimization With ABC Algorithm}
Artificial Bee Colony (ABC) \citep{karaboga2009comparative} is a Swarm Intelligence (SI) algorithm with good exploration and exploitation abilities \citep{vcrepinvsek2013exploration,eiben1998evolutionary}. The main objective of ABC algorithm is to find the food sources by simulating scout, employed, and onlooker bees. The algorithm that we used was presented in \citep{mernik2015clarifying} as "Algorithm 1".

Equation (\ref{13}) is used as an objective function to ensure fairness among users. A constraint is defined in (\ref{14}) to ensure that the transmitter remains at the specified power level. With the constraint shown in (\ref{kanalsir}) and (\ref{15}), more power is allocated to users with weaker channel gain.
\begin{equation}\label{13}
\max _{p_{k} \in p} \quad F=\frac{\left(\sum R_{k}\right)^{2}}{K \sum R_{k}^{2}}
\end{equation}
\begin{equation}\label{14}
\sum_{k=1}^{K}{p_k=P_{max}}
\end{equation}
\begin{equation}\label{kanalsir}
0<h_k<h_{k-1}<p_{k-2}\ldots\ldots..<h_1
\end{equation}
\begin{equation}\label{15}
0<p_1<p_2<p_3\ldots\ldots..<p_k
\end{equation}
In the equations above; $P_{max}$ is the specified power level of the transmitter, $R_k$ is the lower bound expression of VLC channel capacity of user $k$ as defined in (\ref{lowerbound}), $K$ is the total number of users and $F$ is the fairness index.
In order to allocate power efficiently, it is essential to have knowledge of the Channel State Information (CSI) first. \citep{tong2022research} has proposed a compressed sensing-based algorithm for channel estimation in NOMA-VLC systems, aiming to reduce the pilot proportion and enhance communication efficiency. Similarly, \citep{palitharathna2022neural} has introduced a neural network-based channel estimation approach for spatial modulation VLC systems, demonstrating superior accuracy and resolution compared to spline interpolation. As our paper focuses on proposing a fairness-maximized power allocation method for NOMA in VLC, we make the assumption that CSI is known for mobile users.

\begin{table}
\centering
\label{tablo1}
\captionsetup{position=top}
\caption{VLC channel and room parameters}
\begin{tabular}{ll}
\rowcolor{black} \textbf{\textcolor{white}{ Parameter }}    & \textbf{\textcolor{white}{ Value }}  \\
\rowcolor[rgb]{0.8,0.8,0.8} Office dimensions                              & $6m\ \times\ 6m\ \times\ 3m$               \\
PD surface area                                                            & $1cm^2$                       \\
\rowcolor[rgb]{0.8,0.8,0.8} PD refractive index                            & $1.5$                         \\
Optical filter gain                                                        & $1$                           \\
\rowcolor[rgb]{0.8,0.8,0.8} Optical lens gain                              & $3$                           \\
AWGN variance                                                              & $-180$  dBm/Hz \citep{stefan2013area}                           \\
\rowcolor[rgb]{0.8,0.8,0.8} Sum of the user powers
  & $22.5$ W                        \\
LED semi-angle                                                             & ${60}^o$                         \\
\rowcolor[rgb]{0.8,0.8,0.8} Transceiver bandwidth                          & $30$ MHz                      \\
FoV of PD                                                                  & ${60}^o$                       
\end{tabular}
\end{table}

\section{EFOPA}
\label{Numerical Results and Discussions}

\begin{figure}
    \centering
    \subfloat[Simulated communication room\label{area}]{%
        \includegraphics[scale=0.19]{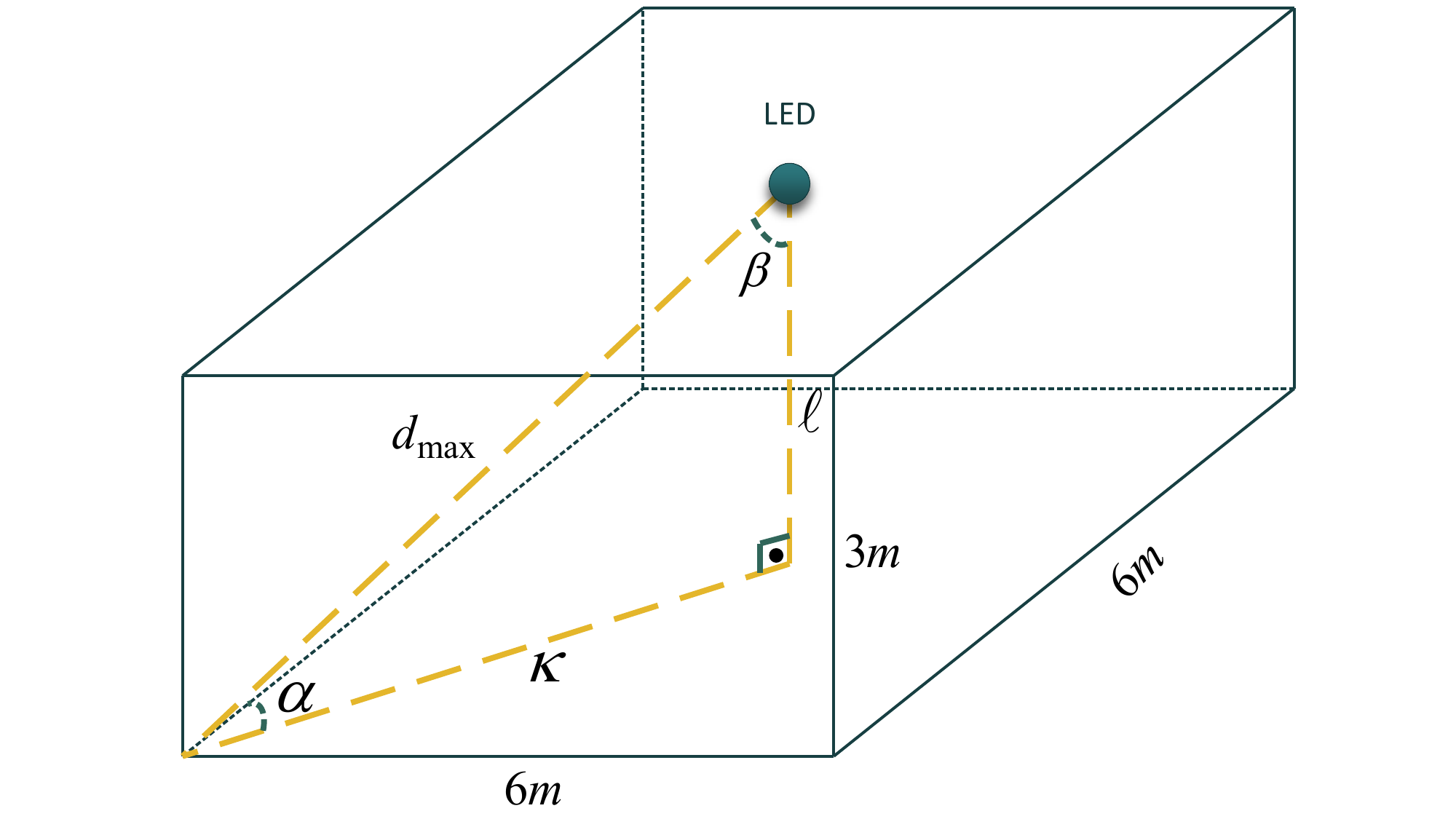}
    } \qquad
    \subfloat[Unique channels in logarithmic scale\label{fig:1544}]{%
        \includegraphics[scale=0.51]{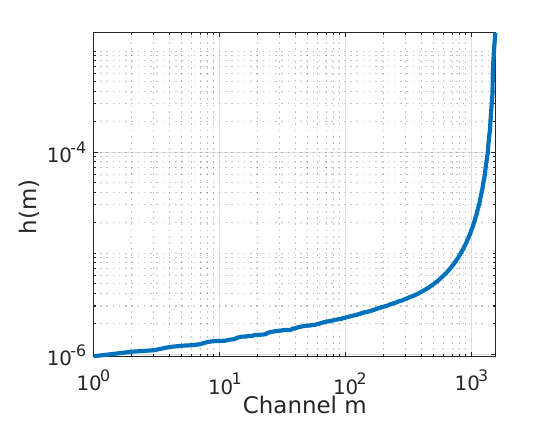}
    }
    \caption{Simulated communication room and unique channels in logarithmic scale}
\end{figure}

The proposed method, EFOPA, is illustrated using a model room depicted in Figure \ref{area}, specifically designed for simulations. Table 1. outlines the values utilized to establish VLC Line of Sight (LoS) channels in this model room. The transmitter LED is strategically placed at the ceiling's center within a simulated empty office room measuring 6x6x3 meters. Employing basic geometry, we calculate the maximum distance ($d_{max}$) between the transmitting LED and the receiving PD as $3\sqrt3$ meters. The distance ($d$) between the transmitter and the receiver is assumed to vary from $0.25$ meters to $3\sqrt3$ meters, with a step size of $0.25$ meters.

Furthermore, we introduce variations in the incidence angle of the PD and the irradiance angle of the LED, with a step size of $5^\circ$, spanning the range $5^\circ \leq \psi, \phi \leq 60^\circ$. This results in 3024 combinations of these variables, generating 1544 unique channels denoted by $h(m); m=1,2,...,1544$. It's noteworthy that any of these channels can occur at any time between the users and the transmitter, as illustrated in Figure \ref{fig:1544}.

To replicate realistic VLC conditions, we carefully select values in Table 1 based on prevalent characteristics observed in typical office environments and standard LED specifications. The office dimensions of $6m \times 6m \times 3m$ represent a standard office size, and the LED semi-angle of ${60}^o$ aligns with widely available LEDs featuring typical beam angles. The PD surface area, refractive index, optical filter gain, optical lens gain, and transceiver bandwidth values are meticulously chosen to ensure an authentic representation of the experimental conditions. 

Additionally, the sum of user powers is set to $22.5$ W, considering the required illumination level for VLC applications. These selected parameter values adhere to established standards and industry practices, forming a realistic foundation for the simulations conducted in this study. It's crucial to emphasize that these values significantly contribute to the reliability and relevance of the experimental results.

In the context of our research, we consider a system with two users denoted as $K$. User 1 is presumed to possess a higher channel gain than User 2. The mean value of channel gains across all combinations of distances ($d$), angles ($\psi$), and phases ($\phi$) is denoted as $h_{0}$. The optimization process involves expressing the channel gain of the first user ($U_{1}$) as proportional to $h_{0}$, with $h_{0}$ determined to be $7.9144\times{10}^{-5}$ through simulation. User 2 ($U_{2}$) is assumed to have a comparatively lower channel gain than User 1.

To streamline the optimization process and reduce the number of variables, we simplify operations by setting the power allocation for User 2 ($p_2$) as $p_{max}-p_1$. Consequently, the upper power limit for $p_1$ adheres to the information theory principle \citep{cai2017modulation} and is constrained to a maximum of $\frac{P_{max}}{2}$. In this particular scenario, the optimization parameters are as follows.
\begin{equation}\label{optp_f}
\max _{p_{1} \in p} \quad F=\frac{\left(\sum R_{k}\right)^{2}}{K \sum R_{k}^{2}}
\end{equation}
\begin{equation}\label{kisit}
s.t.\ \ \ \ p_{1} \leq \frac{P_{\max }}{2}
\end{equation}
The optimization is conducted individually for each distinct channel of User 2 ($U_2$), specifically those with channel gains inferior to $h_{1}$ within the cluster $h(m)$, where $h_{1}$ is held constant. Initially, $h_{1}=2h_0$ is chosen as the starting point. Under this condition, the expression for $h_2$ is as follows.
\begin{equation}
 h_{2}\le h_1,\ \ \ \ h_{2}\in h_k
\end{equation}
The optimization process involved iterating through each $h_{2}$ in the set $h(m)$ that adheres to the condition $h_{2}\le h_1$, seeking $p_1$ values that maximize fairness. The ratio of user channels is denoted as $r=h_{2}/h_{1}$. The discrepancy between this ratio and the power allocation $p_1$ is visually represented in blue in Figure \ref{fig:karsi}.

\begin{table}
\label{tablo2}
\caption{ABC algorithm parameters}
\centering
\begin{tabular}{ll}
\rowcolor{black} \textbf{\textcolor{white}{ Parameter }}    & \textbf{\textcolor{white}{ Value }}  \\
\rowcolor[rgb]{0.8,0.8,0.8} Number of
  variables (D)          & $1$                           \\
Lower bound (lb)                                                & $0$                           \\
\rowcolor[rgb]{0.8,0.8,0.8} Upper bound (ub)                    & $P_{max}/2$                       \\
Population size      (NumberFoods)                                       & $10$                        \\
\rowcolor[rgb]{0.8,0.8,0.8} Maximum Fitness
  Evaluation (MaxFe)& $4000$                        \\
Fitness Function (FF)                                           & (Eq. \ref{optp_f})   \\
\rowcolor[rgb]{0.8,0.8,0.8} Limit& $NumberFoods\times D$                        \\
  
\end{tabular}
\end{table}
Following this, a value of $P_{max}=11.25$ was selected, and the described operations were repeated. Assuming a direct proportionality between LED brightness and input power, the illumination level would be halved based on the calculation for $P_{max}=22.5$. The resulting relationship between $p_1$ and $r$ is visualized in purple in Figure \ref{fig:karsi}. Subsequently, by choosing $h_{1}=h_0$, the initial two steps were reiterated, and the results are displayed in Figure \ref{fig:karsi}. Additionally, outcomes for arbitrary scenarios were presented, specifically for $h_1 = 1.17h_0$ with $P_{max}=22.5W$ and $h_1 = h_0$ with $P_{max}=8.1W$.  The relationship between the allocated powers for $U_1$ and $r$ can be determined using Equation (\ref{expformat}). Subsequently, Equation (\ref{fit1}) is established through curve fitting using the MATLAB tool, focusing on the optimization results with $h_1=2h_0$ and $P_{max}=22.5W$, with exponential curve fitting yielding optimal results for this type of curve. As a result, the allocated powers to $U_1$ can be computed for any given values of $h_1$ and $P_{max}$ using Equation (\ref{anyp}).

\begin{equation}\label{expformat}
Fit\left(r\right)=ae^{br}+ce^{dr}
\end{equation}
\begin{equation}\label{fit1}
{Fit}_1\left(r\right)=0.1018e^{0.01274r}-0.1432e^{-19.04r}
\end{equation}
\begin{equation}\label{nu}
\mu=\frac{h_{R}}{h_{N}}\times\sqrt{\frac{P_{N}}{P_{R}}}
\end{equation}

\begin{equation}\label{anyp}
p_1(r)=\mu\times{Fit}_1(r)
\end{equation} 
$h_{R}$: Reference channel which is used to create fitting function (for example \par \hspace{1mm} $h_R=h_1=2h_0$),\\
$h_{N}$: Any channel for strong user ($U_1$),\\
$P_{R}$: Reference total power which is used to create fitting function (for example \par \hspace{1mm} $P_R=P_{max}=22.5W$),\\
$P_{N}$: Any possible total power desired.

Algorithm (\ref{algo1}) and algorithm (\ref{algo2}) present the pseudo-code of equation generation and usage of EFOPA respectively. \\

\begin{algorithm}[H]

\caption{EFOPA Simplified Algorithm}
\label{algo1}
\textbf{Input:} $\mathcal{H}$ - Set of all possible user channels\;

\textbf{Output:} $\text{CurveEquation}$ - Exponential curve equation\;

\begin{enumerate}
    \item \textbf{Calculate Unique Channels:}\\
    $\mathcal{H}_{\text{unique}} = \text{sort}(\text{unique}(\mathcal{H}))$\;
    
    \item \textbf{Calculate Mean Channel Value (\(h_0\)):}\\
    $h_0 = \frac{1}{length({\mathcal{H}_{\text{unique}}}}) \sum_{h_i \in \mathcal{H}_{\text{unique}}} h_i$\;
    
    \item \textbf{Equalize $h_1$ to $2h_0$:}\\
    $h_1 =  2h_0$\;
    
    \item \textbf{Power Allocation using ABC Algorithm:}\\
    \For{each unique channel \(h_{2}\) in $\mathcal{H}_{\text{unique}}$}{
        \If{$h_{1} < h_2$}{
            Swap $h_{2}$ with $h_1$\;
        }
        $p_1 = \text{ABC}(\text{FF, lb, ub, MaxFe, NumberFoods, limit, } h_{2}, \text{Noise})$\;
        Store $\frac{h_2}{h_{1}}$ and $p_1$\;
    }
    
    \item \textbf{Curve Fitting:}\\
    Import $\frac{h_2}{h_{1}}$ and $p_1$ data\;
    Find $\text{CurveEquation}$\;
    Store $\text{CurveEquation}$\;
\end{enumerate}
\end{algorithm}

\begin{algorithm}[H]
\label{algo2}
\caption{EFOPA Usage Algorithm}

\textbf{Input:} $\text{CurveEquation}$ - Exponential curve equation, $h_k$ - Channel gain of each user\;

\textbf{Output:} Power allocation to users based on channel gains\;

\begin{enumerate}
    \item \textbf{Load Curve Equation:} Load $\text{CurveEquation}$ from Algorithm 1\;
    
    \item \textbf{Get User Channel Gain:} Obtain the channel gain of each user: $h_k$\;

    \item \textbf{Calculate Channel Gain Ratio:} : $r = \frac{h_2}{h_1}$\;

    \item \textbf{Calculate} $\mu$ \textbf{Value:} : $\mu = \frac{2h_0}{h_1}$\;

    \item \textbf{Multiply Curve Equation by} $\mu$: $\text{CurveEquation} = \mu \times \text{CurveEquation}$\;

    \item \textbf{Power Allocation:} Apply the curve equation with $r$: $p_1 = \text{CurveEquation}(r)$\;
\end{enumerate}

\end{algorithm}

\section{Numerical Results}\label{results}

\begin{figure}
\centering
\includegraphics[scale=0.7]{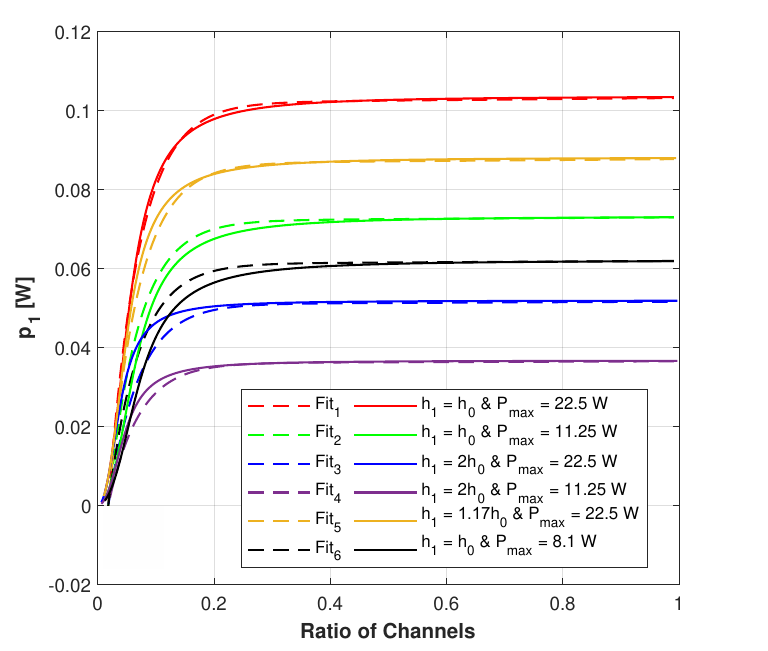}
\caption{Power allocation of User 1 according to the ratio of channels for different total powers and channels.}
\label{fig:karsi}
\end{figure}

In Figure \ref{fig:karsi}, the allocated powers for $U_1$ are compared. The results presented in Figure \ref{fig:karsi} indicate that the proposed power allocation expression (Equation \ref{anyp}) produces outcomes that closely correspond to those obtained through the optimization process.

The proposed EFOPA method streamlines the power allocation process for users, eliminating the need for re-optimization when there are changes in illumination levels or shifts in user locations.

In a specific scenario with $h_1=2h_0$ and $P_{max}=22.5$, we conducted a comparative analysis between our proposed EFOPA method and existing methods, including GRPA, NGDPA, and OMA. Throughout these comparisons, user capacities for both NOMA and OMA were calculated using the Shannon capacity expressions provided in Equations (\ref{shannoncap}) and (\ref{shannoncap2}), respectively.
\begin{equation}\label{shannoncap}
R_{k}=B_{k} \log _{2}\left[1+\frac{h_{k}^{2} p_{k}}{\left(h_{k}^{2} \sum_{l=k+1}^{K} p_{l}+\sigma^{2}\right)}\right]
\end{equation}
\begin{equation}\label{shannoncap2}
R_{k}=\frac{B_{k}}{K} \log _{2}\left[1+\frac{h_{k}^{2} p_{k}}{\sigma^{2}}\right]
\end{equation}

\begin{figure}
    \centering
    \subfloat[EFOPA fairness comparison\label{fig:fairness}]{%
        \includegraphics[scale=0.66]{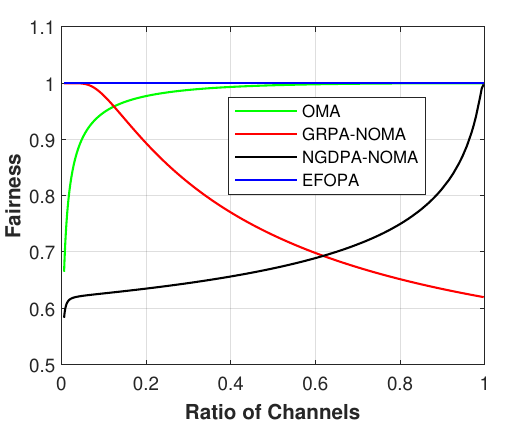}
    } \qquad
    \subfloat[EFOPA sum-rate comparison\label{fig:sumrate}]{%
        \includegraphics[scale=0.66]{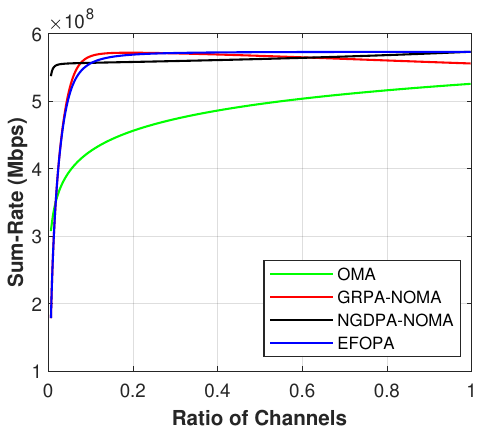}
    }
    \caption{Fairness and sum-rate}
    \label{fig:comp_fairsum}
\end{figure}

Figure \ref{fig:fairness} visually represents the superior fairness achieved by our EFOPA method when compared to alternative methods. GRPA demonstrates poor performance as users' channels become more similar, while NGDPA and OMA methods exhibit inadequate fairness when users' channels differ from each other.

In Figure \ref{fig:sumrate}, EFOPA's superiority over Gain Ratio Power Allocation (GRPA), Normalized Gain Difference Power Allocation (NGDPA), and Orthogonal Multiple Access (OMA) is evident in total capacity comparisons. GRPA and EFOPA yield similar results 50\% of the time, with OMA performing the worst and NGDPA surpassing EFOPA in total capacity only in about 10\% of cases. However, it's crucial to note that higher total capacity in NGDPA doesn't guarantee high fairness. These findings highlight EFOPA's dual advantage: outperforming existing methods in fairness and reducing computational complexity through an empirical equation. EFOPA consistently achieves similar or better sum rates compared to GRPA in almost 50\% of cases, outperforming OMA in nearly 98\% of channel conditions, and surpassing NGDPA in over 90\% of conditions. This underscores EFOPA as a robust and efficient solution for fair power allocation in NOMA-VLC systems.

\begin{figure}%
    \centering
    \includegraphics[scale=0.66]{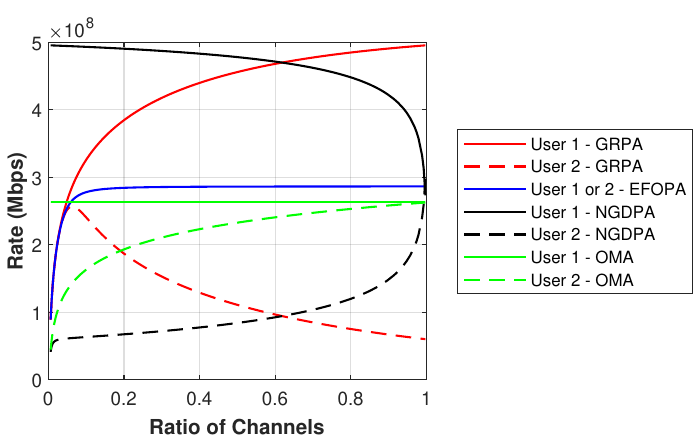}
    \caption{EFOPA user capacity comparison}%
    \label{fig:comp_cap}%
\end{figure}

Figure \ref{fig:comp_cap} shows the user capacities. In NGDPA and GRPA, the channel capacity of one user increases while that of the other decreases as the ratio of the users' channels to each other changes. In the case of OMA, $U_1$ maintains a constant channel capacity, and the channel capacity of $U_2$ approaches that of $U_1$ as the ratio increases. Notably, in the proposed EFOPA method, $U_1$ and $U_2$ consistently exhibit the same channel capacity across all variations of the channel ratio.

\begin{figure}
\label{deneme}
\includegraphics[scale=0.55]{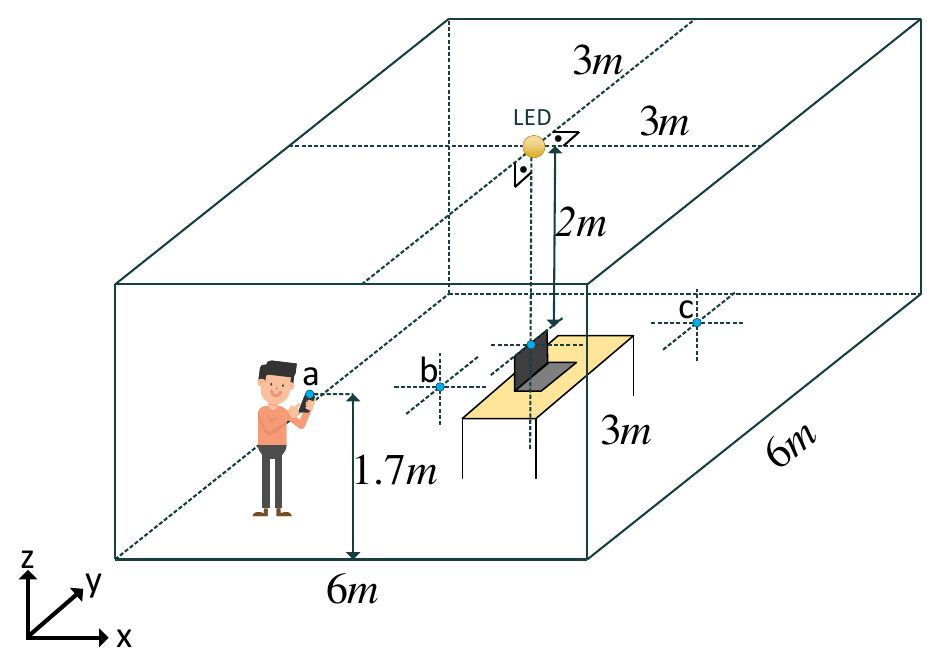}
\centering
\caption{Walking man demonstration}
\label{fig:deneme}
\end{figure}

Figure \ref{fig:deneme} depicts a scenario where a laptop ( $U_1$) remains on a fixed channel with $h_{1}$, while a person holding a mobile phone ( $U_2$) experiences a variable channel $h_{2}$. The individual moves from point $a$ to point $b$ and then to point $c$. In this configuration, the laptop enjoys a stronger channel gain than the mobile phone. For illustrative purposes, the transceiver bandwidth and total power allocations are set to $30MHz$ and $P_{max}=22.5W$, respectively.

Assuming the PD receiver of the laptop is positioned perpendicular to the ceiling at coordinates [3, 3, 1] in the $xyz$ plane, and the distance between the LED transmitter and the laptop is 2 meters. The angle $\phi_{k}$ is set at 0 degrees. It's noteworthy that in this context, $\phi_{k}$ is equivalent to $\psi$ due to the perpendicular alignment of the receiver PD with the ceiling. As a result, we can calculate the channel as:
\begin{equation}
h_{laptop}=h_1=9.5493\times{10}^{-5}
\end{equation}
For the rest points $a=$[2.5, 1.5, 1.7],$b=$[2, 2.5, 1.7],$c=$[4.5, 4, 1.7] the channels for $U_2$ would be;
\begin{equation}
h_{2}\left(a\right)=9.1924\times{10}^{-6},\ \ \  h_{2}\left(b\right)=1.8671\times{10}^{-5},\ \ \  h_{2}\left(c\right)=6.6131\times{10}^{-6}
\end{equation}

Continuing with power allocation for the users and subsequent capacity calculations, we have used $h_0=7.9144\times{10}^{-5}$ as a reference. Therefore, the factor $\mu$ in the equation \ref{nu} is computed as:
\begin{equation}
\mu\ =\ 2h_0/h_{1}=\ 1.6576
\end{equation}
The channel ratio $r$ calculated as:
\begin{equation}
r\left(a\right)=h_{2}/h_{1}=0.0963
\end{equation}
Power allocation is calculated as:
\begin{equation}
p_{1}\left(r\left(a\right)\right)=\mu{Fit}_1\left(r\left(a\right)\right)=p_1(a)=\ 0.0790
\end{equation}
The capacity of $U_1$ is calculated as:
\begin{equation}
R_1(a)=3\times{10}^7\log_2{\left[1+\frac{({9.5493\times{10}^{-5})}^20.0963}{{3\times10}^{-12}}\right]}=237.42 \; Mbps
\end{equation}
The capacity of $U_2$ is calculated as:
\begin{equation}
R_2(a)=3\times{10}^7\log_2{\left[1+\frac{({9.1924\times{10}^{-6})}^2(22.5-0.0963)}{({9.1924\times{10}^{-6})}^20.0963+{3\times10}^{-12}}\right]}=244.61 \; Mbps
\end{equation}
We have replicated the same calculations for points $b$ and $c$, yielding the following results.
\begin{equation}
R_1(b)=246.96Mbps,\ \ \  R_2(b)=235.03\;Mbps
\end{equation}
\begin{equation}
R_1(c)=228.06Mbps,\ \ \  R_2(c)=254.01\;Mbps
\end{equation}
As evident from the calculations, we can achieve power allocations that maximize fairness without the need for re-optimizations. Table 3. presents a comparison of EFOPA to GRPA and NGDPA concerning user speeds, fairness, and total capacity. The EFOPA method consistently achieves maximum fairness across all test positions. In contrast, the fairness index for GRPA varies between $0.78$ and $0.99$. Moreover, NGDPA only manages to provide a fairness index of around $0.52$. We can also see that EFOPA provides higher sum rates than the NGDPA and GRPA. 
\begin{table}
\label{tablo3}
\caption{EFOPA performance comparison}
\begin{tabular}{l}
\includegraphics[scale=0.72]{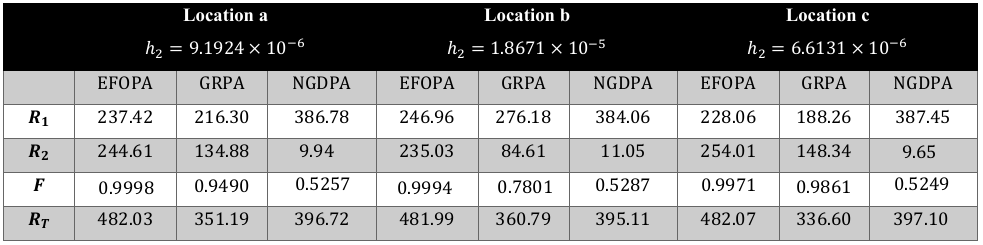}\end{tabular}
\end{table}                

\section{Conclusion}\label{Conclusion}
This article introduces a novel approach, the EFOPA method, for achieving maximum fairness in PD-NOMA VLC systems. The proposed method ensures optimal fairness across various channel conditions and different levels of illumination without the need for re-optimization, offering a streamlined communication system even as user locations or illumination levels change. This results in reduced computational complexity requirements for maintaining fairness in communication systems. Comparative analyses with existing methods, such as GRPA and NGDPA, highlight the superiority of EFOPA in terms of fairness. EFOPA consistently delivers maximum fairness performance in PD-NOMA VLC systems, contrasting with the instability observed in GRPA and NGDPA as the ratio of users' channels fluctuates. Future work will delve into user pairing schemes employing the EFOPA method, along with an exploration of scenarios involving multi-color LED transmitters in NOMA-VLC.


\end{document}